\documentclass[]{spie}  %>>> use for US letter paper
\usepackage{amsmath,amsfonts,amssymb}
\usepackage{graphicx}
\usepackage{physics}
\usepackage{multirow}
\usepackage{multicol}
\usepackage{booktabs}
\usepackage{adjustbox}
\usepackage[colorlinks=true, allcolors=blue]{hyperref}

\title{High-performance Data Management for Whole Slide Image Analysis in Digital Pathology}

\author[a]{Haoju Leng}
\author[a]{Ruining Deng}
\author[b]{Shunxing Bao}
\author[a]{Dazheng Fang}
\author[c]{Bryan A. Millis}
\author[d]{Yucheng Tang}
\author[e]{Haichun Yang}
\author[g]{Xiao Wang}
\author[h]{Yifan Peng}
\author[f]{Lipeng Wan}
\author[a,b,e]{Yuankai Huo}

\affil[a]{Department of Computer Science, Vanderbilt University, Nashville, TN, USA}
\affil[b]{Department of Electrical and Computer Engineering, Vanderbilt University Medical Center, Nashville, TN, USA}
\affil[c]{Department of Biomedical Engineering, Vanderbilt University Medical Center, Nashville, TN, USA}
\affil[d]{NVIDIA Corporation, Bethesda, MD, USA}
\affil[e]{Department of Pathology, Microbiology and Immunology, Vanderbilt University Medical Center, Nashville, TN, USA}
\affil[f]{Department of Computer Science, Georgia State University, Atlanta, GA, USA}
\affil[g]{Oak Ridge National Laboratory, Oak Ridge, TN, USA}
\affil[h]{Department of Population Health Sciences, Weill Cornell Medicine, New York, NY, USA}

\authorinfo{Corresponding author: Yuankai Huo: E-mail: yuankai.huo@vanderbilt.edu}

\usepackage{silence}
\begin{document} 
\maketitle

\begin{abstract}
When dealing with giga-pixel digital pathology in whole-slide imaging, a notable proportion of data records holds relevance during each analysis operation. For instance, when deploying an image analysis algorithm on whole-slide images (WSI), the computational bottleneck often lies in the input-output (I/O) system. This is particularly notable as patch-level processing introduces a considerable I/O load onto the computer system. However, this data management process could be further paralleled, given the typical independence of patch-level image processes across different patches. This paper details our endeavors in tackling this data access challenge by implementing the Adaptable IO System version 2 (ADIOS2). Our focus has been constructing and releasing a digital pathology-centric pipeline using ADIOS2, which facilitates streamlined data management across WSIs. Additionally, we've developed strategies aimed at curtailing data retrieval times. The performance evaluation encompasses two key scenarios: (1) a pure CPU-based image analysis scenario (``CPU scenario"), and (2) a GPU-based deep learning framework scenario (``GPU scenario"). Our findings reveal noteworthy outcomes. Under the CPU scenario, ADIOS2 showcases an impressive two-fold speed-up compared to the brute-force approach. In the GPU scenario, its performance stands on par with the cutting-edge GPU I/O acceleration framework, NVIDIA Magnum IO GPU Direct Storage (GDS). From what we know, this appears to be among the initial instances, if any, of utilizing ADIOS2 within the field of digital pathology. The source code has been made publicly available at \url{https://github.com/hrlblab/adios}.
\end{abstract}


\section{INTRODUCTION}
%\section{Description of purpose}
\label{sec:intro}  % \label{} allows reference to this section
The introduction of the Whole Slide Image (WSI) approach and other recent advancements in digital pathology has significantly transformed the field. Digital pathology has revolutionized the way pathologists work by providing the opportunity to analyze slides remotely through monitors, eliminating the need for traditional local microscopes~\cite{bengtsson2017computer}\cite{marti2021digital}\cite{gomes2021building}. Furthermore, it has opened new avenues for computer-assisted quantification, empowering pathologists with advanced tools for image analysis~\cite{madabhushi2016image}\cite{eriksen2017computer}\cite{kothari2013pathology}. However, the large size of these images presents significant computational challenges, particularly in the context of image analysis pipelines. One of the major bottlenecks in this process lies in the I/O system, as patch-level processing requires frequent data access, leading to considerable I/O load on the computer system~\cite{deroulers2013analyzing}.

To address this data access and management challenge across WSIs, we present our efforts in implementing the Adaptable IO System version 2 (ADIOS2), a powerful and flexible framework for handling large-scale scientific data~\cite{godoy2020adios}. We have been dedicated to building and launching a digital pathology-centric pipeline powered by ADIOS2, enabling us to efficiently manage data across WSIs. By efficiently handling large-scale data and enabling parallel data access, ADIOS2 opens new possibilities for accelerating image analysis pipelines.

% CPU scenario要说是image analysis吗？
In this paper, we conducted a comprehensive performance evaluation by considering two key scenarios: (1) the ``CPU scenario", where image analysis is performed using a pure CPU-based approach, and (2) the ``GPU scenario", where we leverage the power of deep learning frameworks on the GPU. Our findings reveal noteworthy results. Under the CPU scenario, ADIOS2 showed an impressive two-fold speed-up in I/O operations compared to the brute-force approach, demonstrating its efficiency in handling large-scale data. In the GPU scenario, ADIOS2's performance matches the state-of-the-art GPU I/O acceleration framework, NVIDIA Magnum IO GPUDirect Storage (GDS)~\cite{nvidia-gpudirect-storage}. Based on our knowledge, this is one of the first, if any, applications of ADIOS2 in digital pathology. The source code has been made publicly available at \url{https://github.com/hrlblab/adios}.

% \begin{figure}[t]
% \begin{center}
% \includegraphics[width=0.75\linewidth]{fig/fig_1.pdf}
% \end{center}
%    \caption{This figure presents an overview of the proposed pipeline. The input is the raw WSIs, while the output is the completely labeled tissue segmentation result. The domain impacts are to 1) enable slide-wise multi-tissue segmentation for WSIs via one command for non-technical users and to 2) achieve better segmentation quality with less time.}
% \label{fig:problem}
%  \end{figure}

\section{Method}
\begin{figure}[t]
\begin{center}
\includegraphics[width=1.0\linewidth]{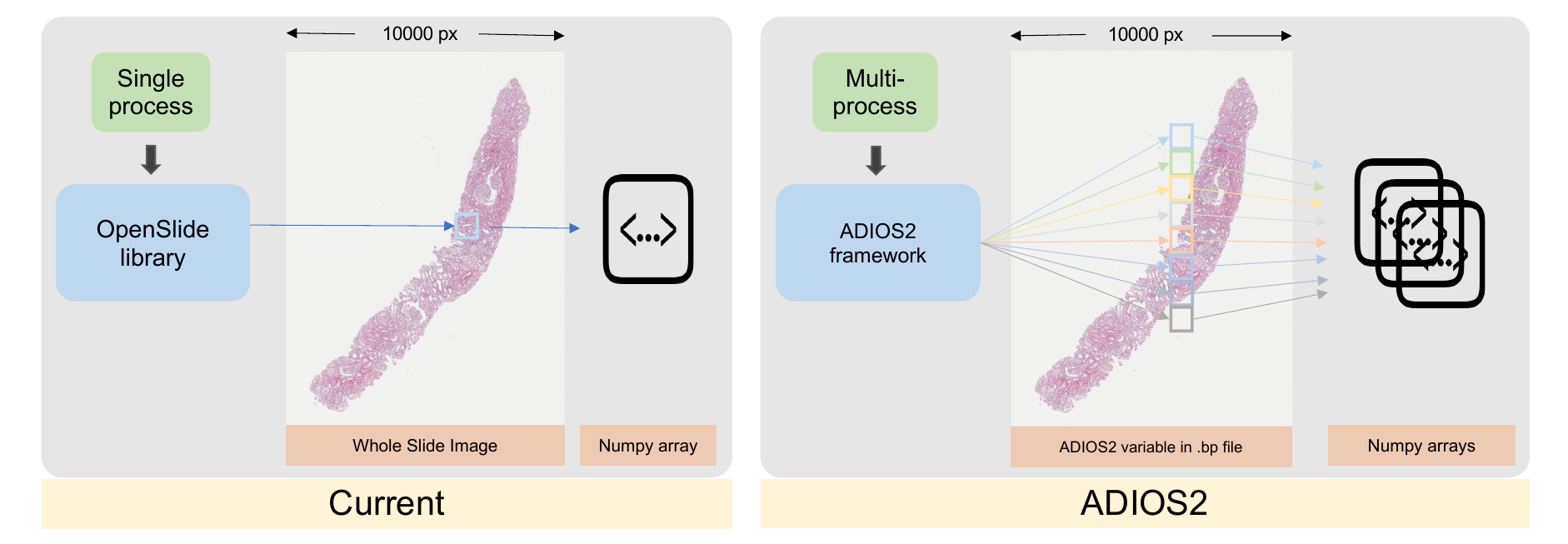}
\end{center}
   \caption{The comparison between the current prevalent I/O methods and the ADIOS2 framework approach in digital pathology image analysis pipelines.}
\label{fig:idea}
 \end{figure}

\begin{figure}[t]
\begin{center}
\includegraphics[width=1.0\linewidth]{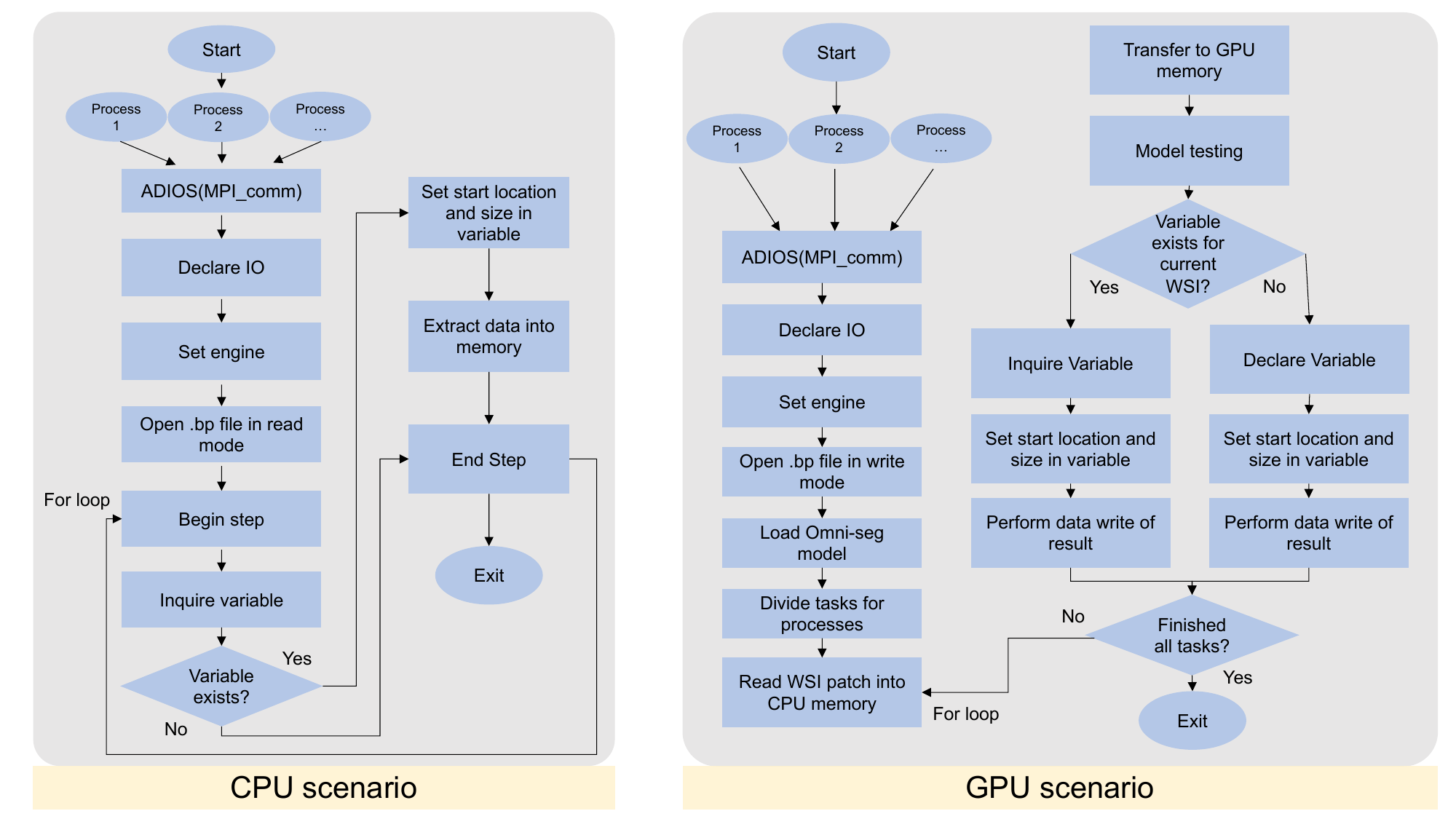}
\end{center}
   \caption{The ADIOS2 framework in the CPU scenario (left panel) and GPU scenario (right panel) experiments.}
\label{fig:flowchart}
 \end{figure}

ADIOS2 utilizes the Message Passing Interface (MPI)~\cite{walker1996mpi} for parallel I/O operations. It can read and write self-describing data in binary-packed format (.bp). In the .bp file, data is written to unique user-specified variables, which are classified by data types and shapes. Data can also be extracted from specific variables~\cite{godoy2020adios}. The comparison between prevalent I/O methods and the ADIOS2 framework is demonstrated in Fig.\ref{fig:idea}. A flow chart showing the steps of the ADIOS2 framework in the CPU and GPU scenarios is presented in Fig.\ref{fig:flowchart}. The performance evaluation has two parts: (1) a pure CPU-based image analysis scenario (``CPU scenario"); and (2) a GPU-based deep learning framework scenario (``GPU scenario").

\subsection{CPU scenario}
ADIOS2 reading performance is evaluated by retrieving data from the .bp file, which was converted from WSIs of 40$\times$ magnification using ADIOS2. The program iterated through all variables, and $512\times512$ patch data was selected from the variables iteratively. Then, the data was assigned to a numpy array, mimicking accessibility to an image analysis pipeline.

To retrieve the array of the entire WSI, saved as a numpy data format (.npy) file, we used a single process to load the .npy file directly into CPU memory and then extract $512\times512$ patch arrays sequentially from the large numpy array. For the small patch arrays saved as .npy files, we used multiple processes to load .npy files into numpy arrays in parallel.

% \subsubsection{Writing performance evaluation}
% To evaluate performance using multi-process ADIOS2 framework, the variable was defined as the equivalent shape of the WSI. Patch of size $512\times512$ pixels was extracted from the WSI as a numpy array, which mimics the patch extraction process before segmentation. The array was then wrote to the corresponding location, which is identical as the location on WSI, of the variable. We continued this process until it covered the whole WSI region. Each process was assigned to perform writing operation to ensure the patches were only wrote once. We measured time of writing data from CPU memory into disk, ignoring time used to load data from WSI into CPU memory.

% To compare with the ADIOS2 performance, we used two methods separately to write array data from CPU memory to numpy array data format (.npy) file. One was using single process to extract entire WSI as one numpy array and then it was saved as .npy data format. The other was using multi-process to extract $512\times512$ patches from the WSIs and saved as .npy data format files respectively. The patches were assigned to a unique process to avoid repeated writing of the patch. 

% \subsubsection{Reading performance evaluation}

\subsection{GPU scenario}
\begin{figure}[t]
\begin{center}
\includegraphics[width=1\linewidth]{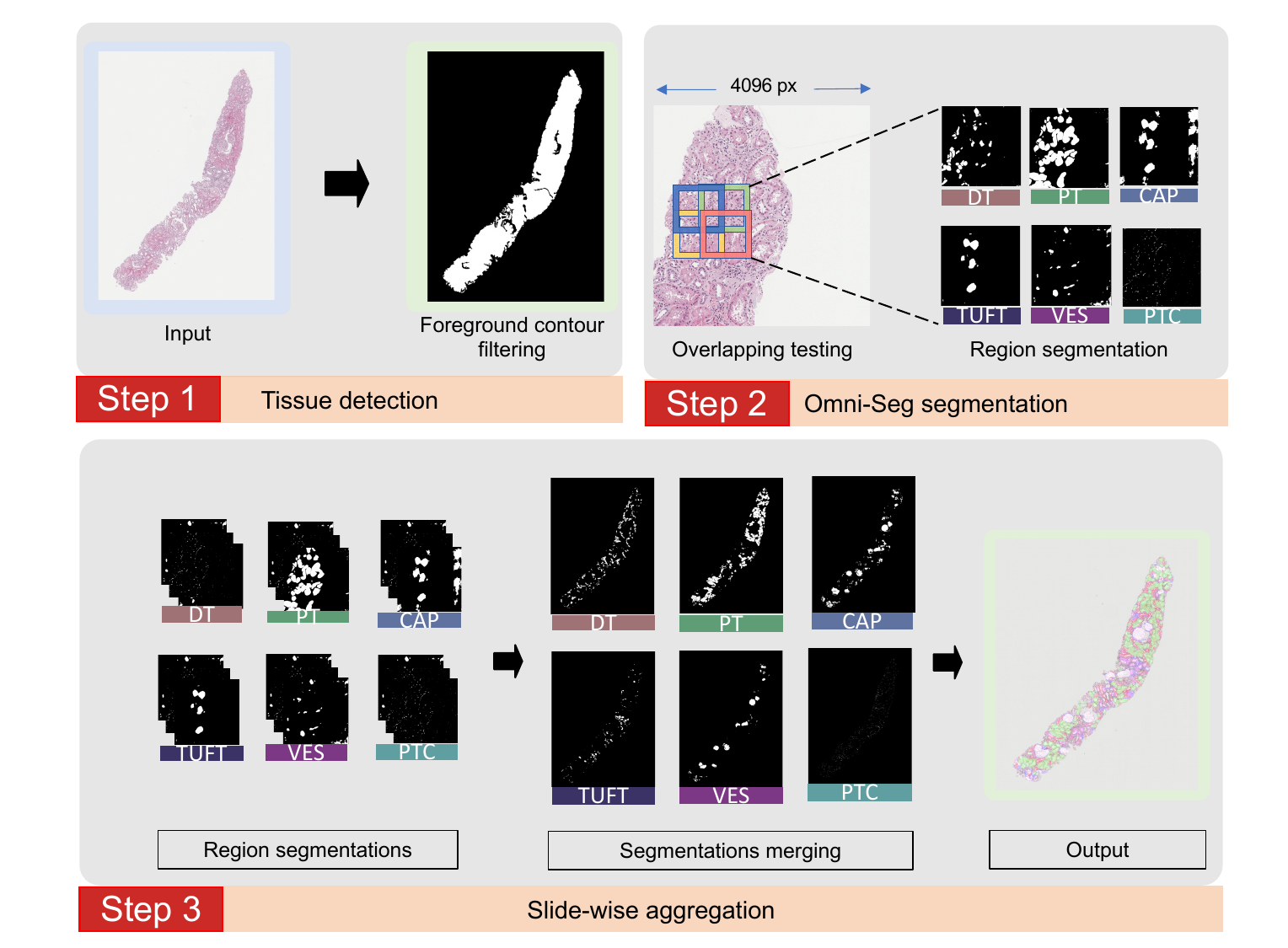}
\end{center}
   \caption{\textbf{Accelerated+ pipeline.} The figure illustrates the comprehensive framework of the proposed pipeline. Step 1 involves tissue detection of the input, while Step 2 focuses on Omni-Seg segmentation of patches derived from Step 1. Finally, Step 3 entails aggregating the results obtained from Step 2. Both Step 1 and Step 3 benefit from GPU acceleration, which enhances the overall efficiency of the process.}
\label{fig:pipeline}
 \end{figure}

The Accelerated+ Omni-seg pipeline is an advanced image segmentation framework that employs an efficient multi-label segmentation technique for pathology quantification~\cite{leng2023accelerated}. The overall framework of the Accelerated+ Omni-seg pipeline is demonstrated in Fig.\ref{fig:pipeline}. We used three approaches on the same Accelerated+ Omni-seg image segmentation pipeline: (1) original approach; (2) multi-process ADIOS2 framework approach; and (3) NVIDIA Magnum IO GPUDirect Storage (GDS) approach.

\subsubsection{Single-process original approach}
The pipeline was the same as the first two steps of the Accelerated+ Omni-seg pipeline. Its first step was to read the input in .PNG data format into GPU memory, add paddings to the image array, and extract $4096\times4096$ patches from it to save as .npy data format. The second step was to apply the pre-trained Omni-Seg model to selected patches, which were retrieved from .npy data format files saved from the first step. The results were moved from GPU memory to CPU memory and then saved as .npy data format. The start time and end time were recorded to calculate the total time of the experiment.

\subsubsection{Multi-process ADIOS2 framework approach}
We modified the Accelerated+ Omni-seg pipeline to accept WSIs as input without converting them into PNG data format first. The WSIs were divided into regions, each assigned to a specific process for preprocessing and image segmentation. For each region, $4096\times4096$ patches were extracted from the WSI into GPU memory. Paddings were added to the patch arrays based on their location in the WSI. After preprocessing, we applied the pre-trained Omni-seg model for segmentation. The results were moved from GPU to CPU and then saved as a .bp data format. We recorded the total time of the pipeline execution.

\subsubsection{Single-process NVIDIA Magnum IO GPUDirect Storage approach}
GDS creates a direct and efficient data pathway between GPU memory and local storage, facilitating smooth data movement to and from GPU memory without relying on the CPU. This ensures optimized performance and minimized CPU overhead~\cite{nvidia-gpudirect-storage}.

The pipeline was modified to accept WSIs as the image input. $4096\times4096$ patches were extracted from the WSI into GPU memory, and then we added paddings to the patches based on their location in the WSI. The pre-trained Omni-seg model was then applied to the patches. The result was saved directly from GPU memory to disk as cuFile data format~\cite{cuFILE-nvidia-gpudirect-storage}. The process was iterated until it covered the entire WSI. The total time needed for the whole program is measured.

\section{Data and Experimental design}
%\section{Data}
\subsection{Data}
For our experimental evaluation, we selected a diverse set of 100 kidney Whole Slide Images (WSIs) from the Kidney Tissue Atlas (KPMP) Datasets. These WSIs were acquired at the original pixel resolution of 0.25 Micron, ensuring a high-resolution representation of kidney tissue. To capture a comprehensive view of different tissue structures, the slides were stained using a variety of techniques, including Hematoxylin and Eosin (H\&E), Periodic-acid-Schiff (PAS), Silver (SIL), Toluidine Blue (TOL), and Trichrome (TRI) staining.

%keep this or not?
% The regions of interests - including glomerular tuft (TUFT), glomerular unit (CAP), proximal tubular (PT), distal tubular (DT), peritubular capillaries (PTC), and arteries (VES) - were manually annotated by the ImageScope software for the GPU scenario.

\subsection{Experimental Design}
\subsubsection{CPU scenario}
Our experimental setup involved a total of 100 WSIs, which were divided into five runs. In each run, we processed 20 WSIs to gather comprehensive results. We used the openslide library~\cite{goode2013openslide} for WSI data extraction to generate testing files. The patch size was $512\times512$ pixels across all three methods. Eight processes were leveraged for the ADIOS2 framework. The ADIOS2 engine was set to BP5 and configured in deferred mode. A single process was utilized to load the WSI-wise .npy file into CPU memory, while eight processes were used for multi-process reading of the patch-wise .npy files from the disk into CPU memory. We used NVIDIA RTX A5000 24G in our experiments. The disk capacity is 5.0 TB, and the CPU memory capacity is 32G.
%We started to measure time usage after fully loading WSI data into the CPU memory for all three approaches. 
\subsubsection{GPU scenario}
The pipeline used for evaluating the single-process original approach was mostly the same as the first two steps of the Accelerated+ Omni-seg pipeline. For the multi-process ADIOS2 framework approach, we used 3 processes to run the pipeline. Each process was assigned an equal number of patches if they could be divided equally. Otherwise, any extra patches were assigned to the last process to ensure that all patches were processed. The implementation was modified to support WSIs as input. As for the single-process GDS approach, the implementation was changed to support WSIs input and utilized GDS for data writing, in comparison to the original pipeline. All other image-analysis-related parameters were kept the same as the parameters of the Accelerated+ Omni-seg pipeline. We used NVIDIA RTX A5000 24G in our experiments. The disk capacity is 5.0 TB, and the CPU memory capacity is 32G.
%We recorded the start time and end time of the pipeline to calculate the total time of the experiment. 

\section{Results}
We conducted 100 repeated experiments using the same 100 kidney WSIs in the CPU scenario. The reading performance is illustrated in Table.\ref{table:CPU scenario reading}. It indicates that the ADIOS2 method required less time for each data split compared to reading WSI-wise and patch-wise .npy files.

For the GPU scenario, we performed 10 repeated experiments using 10 kidney WSIs with the PAS stain. The performance evaluation is presented in Table.\ref{table:GPU scenario}. The table shows that the NVIDIA GDS method required less time than the original and ADIOS2 methods. Interestingly, the ADIOS2 and the original methods showed similar average processing times for each WSI.

\begin{figure}
\begin{center}
\includegraphics[width=1.0\linewidth]{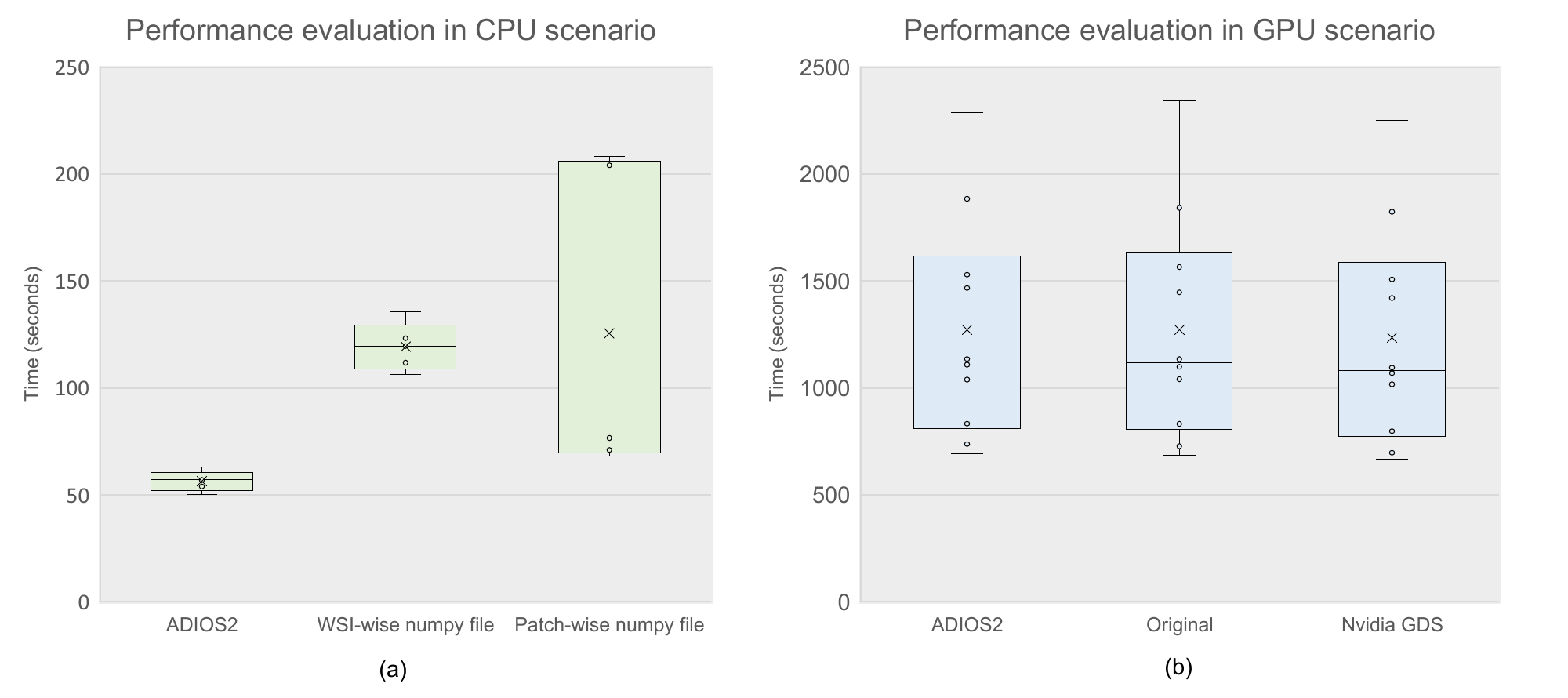}
\end{center}
   \caption{This figure shows the processing time results in CPU and GPU scenarios (a) The total time needed for CPU scenario experiments (b) The total time needed for GPU scenario experiments.}
\label{fig:speed}
 \end{figure}

Fig.\ref{fig:speed} also depicts the performance evaluation in both the CPU and GPU scenarios. As shown in Fig.\ref{fig:speed}, ADIOS2 achieved the lowest median total processing time in the CPU scenario and comparable median total processing time to the original approach in the GPU scenario.

\begin{table}[h]
\caption{Quantitative result of reading performance in CPU scenario.}
\centering
\begin{tabular}{l|ccc}
\toprule
% \noalign{\smallskip}
 methods & Total time(s)& Mean time(s)/WSI & Standard deviation(s)\\
\midrule
 WSI-wise numpy file  & 596.4 & 119.3 & 11.3\\ 
 Patch-wise numpy file & 628.0 & 125.6 & 73.7\\
 ADIOS2 & \textbf{282.0} & \textbf{56.4} & \textbf{4.8} \\
\bottomrule
\end{tabular}
\label{table:CPU scenario reading}
\end{table}

\begin{table}[h]
\caption{Quantitative result of baseline methods in GPU scenario.}
\centering
\begin{tabular}{l|ccc}
\toprule
% \noalign{\smallskip}
 methods & Total time(s)& Mean time(s)/WSI & Standard deviation(s)\\
\midrule
 Original & 12720.9 & 1272.1 & 530.3 \\
 NVIDIA GDS & \textbf{12348.2} & \textbf{1234.8} & \textbf{515.2}\\
 ADIOS2  & 12719.3 & 1271.9 & 519.4 \\
\bottomrule
\end{tabular}
\label{table:GPU scenario}
\end{table}

% \section{New or breakthrough work to be presented}
% In this study, we propose an improved version of the Omni-Seg pipeline in order to reduce the repetitive computing process. The proposed pipeline achieves better model performance and a faster speed of prediction. The entire process can be run in one line of command with the help of a Docker. 

\section{Conclusion}
Based on our results, the ADIOS2 framework demonstrates a reduced time cost for large-scale I/O operations in the pure CPU scenario. In the GPU scenario, where the usage of GPU already reduces substantial processing time in the image analysis pipeline, the I/O time reduction in the ADIOS2 framework is not as pronounced compared to the original method. However, it still achieved comparable performance compared with the WSI-optimized NVIDIA GDS algorithms.

\section{ACKNOWLEDGMENTS}       
This work has not been submitted for publication or presentation elsewhere. This work is supported in part by NIH R01DK135597(Huo), DoD HT9425-23-1-0003(HCY), NIH NIDDK DK56942(ABF) and NSF CAREER Award 2145640(YP). This manuscript has been co-authored by UT-Battelle, LLC, under contract DE-AC05-00OR22725 with the US Department of Energy (DOE). The publisher acknowledges the US government license to provide public access under the DOE Public Access Plan (http://energy.gov/downloads/doe-public-access-plan).

% References
\bibliography{main} % bibliography data in report.bib
\bibliographystyle{spiebib} % makes bibtex use spiebib.bst

\end{document}